\documentclass[12pt]{book}

\usepackage[dvips]{graphicx,color}
\usepackage{makeidx,tsukuba}
\makeauthorindex
\makeindex
\def\thefootnote{\fnsymbol{footnote}}

\begin{document}

\BookTitle{\itshape The 28th International Cosmic Ray Conference}
\CopyRight{\copyright 2003 by Universal Academy Press, Inc.}
%\tableofcontents
\pagenumbering{arabic}

\chapter{Calibration of the Pierre Auger fluorescence detector}

\author{%
M. D. Roberts$^1$ for the Auger Collaboration$^2$\\
{\it (1) Dept. of Physics, University of New Mexico, 
Albuquerque, NM 87131-1156, USA\\
(2) Observatorio Pierre Auger, Av. San Martin Norte 304, (5613) 
Malargue, Argentina} \\
}%% end of author

\section*{Abstract}
The absolute calibration of an air fluorescence detector (FD) is an  
important element in correctly determining the energy of detected  
cosmic rays. The absolute calibration relates the flux of photons of a  
given wavelength at the detector aperture to the electronic  signal  
recorded by the FD data acquisition system.  For the Auger FDs, the  
primary absolute calibration method uses a diffusive surface which is  
placed in front of a telescope aperture to uniformly illuminate the  
telescope field of view with a known light signal.  This  
single-wavelength measurement (375 nm) will be made at intervals of  
several months until the stability of the telescopes is determined. The 
relative wavelength dependence of the calibration is determined through
independent measurements.  
The error in absolute calibration at a single wavelength  
is estimated to be less than 10\%. Two other absolute calibration  
methods are used to provide an  independent verification of the primary  
measurement.  
The stability of the calibration with time is monitored nightly by 
a relative calibration system.
In this paper we will provide descriptions of the  
absolute and relative  calibration  methods used by the Auger air  
fluorescence observatory. Results from the calibration of the Auger  
Engineering Array telescopes will also be presented.

\section{Introduction}
The absolute calibration of the Auger fluorescence detectors
is determined by the wavelength dependent
photon detection efficiency of the FD optical system. 
The components of this system are described in detail in 
[3]. The interaction between these components makes it difficult to determine
the calibration of the system by measuring the efficiencies of individual
components.
Two ``end to end'' methods, described below, are used  
to determine the absolute calibration. Also described are the relative
calibrations which monitor the time and wavelength variability
in the calibration.

\section{The ``Drum'' calibration}
The drum calibration is designed to provide an end to end calibration 
of the FD using an illuminated surface covering the entrance aperture of
an FD telescope. Ideally, each part of this surface would be uniformly
illuminated and would emit as a Lambertian source over the angular 
range viewed by the camera (21$^{\circ}$).
The mechanical structure of the drum is made from a lightweight aluminum 
frame 2.8 m in diameter and 1.5 m deep. The frame supports the illuminated
front surface of the drum, the reflective back and side surfaces of the
drum and the light source. The frame has coupling points that
accurately position and hold the drum over the telescope aperture during 
calibration measurements.
The illuminated front surface of the drum is made from 0.38 mm 
thick Teflon and has a diameter of 2.5 m. 
The light source consists of 2 pulsed UV LEDs ($375 \pm 12$ nm). The LEDs are
embedded in a Teflon cylinder that sits on top of a 15 cm disk of diffusively
reflective Tyvec. A UV enhanced silicon detector at the top of the
Teflon cylinder monitors the relative
light output of the LEDs and also prevents the direct illumination of
the drum by light from the Teflon cylinder.  
The Tyvec disk is held by struts on the inside of the drum
near to the front surface but facing to the back of the drum. The light from
the disk illuminates the back and sides of the drum, also made of Tyvec, which
diffusively reflect light on to the front surface of the drum.

The angular and spatial uniformity of the drum front surface were measured 
by powering the UV LEDs continuously and measuring the surface 
brightness with a CCD camera. The illumination of the
area of the surface corresponding to
the telescope aperture was found to be spatially uniform
to $\pm$1\%. The angular dependence of the surface brightness followed the
expected cosine (viewing angle) dependence to within $\pm$2\%.
The reference standard for
the calibration of the drum was a UV-enhanced Si detector 
(UDT sensors UV100 photodiode)
calibrated at the National Institute for Standards and Technology (NIST).
This device was not sensitive enough to detect the pulsed LED signal from the
drum so it was used to calibrate a light source which in turn was used to
calibrate a photomultiplier tube (PMT) and its readout ADC. 
This PMT/ADC  was used to measure 
the light flux from the drum front surface.

A drum calibration of telescopes 4 and 5 of the 
Los Leones Engineering Array
FD was made during February/March of 2002.
The calibration of telescope 4 was found to be 
4.0$\pm$0.3 photons at the front of the aperture per FADC count.
For telescope 5 the calibration was 
found to be 5.6$\pm$0.4 photons at the front
of the aperture per FADC count. A complete description of the prototype drum
and of the calibration of the Los Leones Engineering Array telescopes can
be found in [1]. 
A number of new drums are being built to calibrate the FDs of the
full Auger Observatory. The physical construction of the drums
 has been changed to provide
a more robust mechanical structure that can be transported easily between 
sites within the Auger Observatory.
The drums will be used to calibrate the telescopes of the Auger FD 
observatories as they come on line starting mid-2003. Initially, each 
telescope of each observatory will be calibrated every 2-3 months. The
time interval between drum calibrations will be adjusted according 
to the stability of the calibration.
There is currently a development effort underway to make the drums a
multi-wavelength calibration source. The LED light sources will be replaced
with optical fibers which bring light from a monochromator.  

\section{Atmospheric laser scattering calibration}

Artificial tracks, generated from a 355 nm laser located 3-4 km from the FD, are
used to confirm the FD calibration obtained by the drum. 
If the energy and polarization state of the laser beam 
are well known, Rayleigh scattering from the molecular atmosphere
provides an accurately calculable flux of photons arriving at a telescope
aperture from each part of the 
track. An advantage of this calibration method over the drum calibration
is that it produces a track image that is very similar to the cosmic ray
tracks measured by the FD.
A description of a laser system used for this type of measurement 
can be found in [4] .
A set of laser measurements was made on 
March $17^{th}$ 2002 to calibrate telescope 4 of the Engineering Array FD at
Los Leones. 
An analysis of these laser shots gave a calibration
for telescope 4 of 4.1$\pm$0.5 photons at the front of the 
aperture per FADC count
at 355 nm. The two main sources of uncertainty
are the measurement of the absolute energy of the laser shots 
and the contribution of aerosol scattering
and extinction to the measurement.
The absolute calibration of the laser energy
, traceable to NIST standards, is better than 10\%. 
The calibration value found by the laser method agrees very well with
that of the drum measurement made two weeks earlier. Although the drum
and laser calibrations were made at different wavelengths, the wavelength
dependence of the FD filters and that of the PMTs cancel almost
exactly between 375 nm and 355 nm. A full description
of the March $17^{th}$ laser calibration can be found 
in [4].

A new roving laser system is being built for use with the Auger Observatory
which will have smaller uncertainty in laser energy than the 
prototype laser system.
It is planned to use the laser beam calibration as an occasional cross-check
of the drum calibration. If serious discrepancies are found between the
drum and laser calibrations a more comprehensive program of laser
calibrations will be undertaken.

\section{Monitoring of the wavelength and time dependence of the calibration}

The relative wavelength dependence of the calibration will be 
determined through ``piece by piece'' measurements of samples from   
components of the FD telescopes. 
Using a ray tracing simulation of the FD optics these measurements 
can also be used to estimate
the absolute calibration of the FD. Preliminary measurements have 
shown that this calibration is consistent with 
measurements made by the drum and laser methods, but
the uncertainty is large (~20\%).

 To monitor changes in the FD response, a relative
calibration system has been implemented. This relative calibration system
uses 3 xenon flash light sources located at each FD site. Light from
each source (denoted A,B or C) is distributed to the six telescopes  
through 
optical fibers.
Light from the A source goes to a diffuser at the center
of each mirror and directly illuminates the camera. The A source includes a
broadband Johnson-U filter and a filter wheel with
5 different neutral density filters. 
The wavelength dependence of the Johnson-U filter transmission is roughly
matched to that of the FD.
The purpose of the A source is to 
measure the stability and linearity of the camera system.  
Light from the B source is reflected from the mirror on to the
camera to monitor the combined stability of the mirror reflectivity and 
camera gain. The B source also includes a Johnson-U filter.
Light from the C source illuminates diffusively reflective 
Tyvec targets on the inside of the 
telescope doors. The light from the Tyvec goes through the aperture to
the mirror and on to the camera. The C light source includes 5 narrow
band interference filters at wavelengths of 330 nm, 350 nm, 370 nm, 390 nm and
410 nm. This light source monitors the end-to-end
stability of the telescopes at all 5 wavelengths.
The A,B and C calibrations are made at least once for each night
that the FD takes data. Calibration data taken during the running of
the FD Engineering Array has shown that the overall stability of the camera
gain, as well as pixel to pixel variation in relative gain, is better than
5\%. Further details of the relative wavelength and time 
dependent calibrations can be found in [2]. 

\section{Conclusion}

The absolute calibration of the Engineering Array FD telescopes
has been measured using three independent techniques: drum calibration,
laser calibration and piece by piece calibration. Within the 
estimated uncertainties the calibration of FD telescopes measured by 
all three techniques agree.

\vspace{\baselineskip}

\re
1.\ Brack, J., Meyhandan, R. and Hofman, G. 2002, Auger GAP note GAP-2002-033
\re
2.\ Matthews, J. A. J. 2002, in Proc. SPIE conference on Astronomical 
Telescopes and Instrumentation, 22-28 August
\re
3.\ Matthiae G. 2001, in Proc. Hamburg ICRC, Vol. 2 p 733 
\re
4.\ Roberts, M., Sommers, P. and Fick, B. 2003, Auger GAP note GAP-2003-010 

\endofpaper
\end{document}